# Diffusion of oxygen vacancies formed at the anatase (101) surface: An activation-relaxation technique study

Jeffrey Roshan De Lile [1]* and Normand Mousseau [1,2]*

[1] Department of Physical Engineering, Polytechnique Montréal, Case postal 6079, Station Centre-Ville, Montréal, QC H3C 3A7, Canada.
[2] Succursale Centre-Ville, Département de Physique, Université de Montréal, Case Postale 6128, Montréal, QC H3C 3J7, Canada.

**ABSTRACT**

TiO$_2$ is a technologically important material. In particular, its anatase polymorph plays a major role in photocatalysis, which can also accommodate charged and neutral vacancies. There is, however, scant theoretical work on the vacancy charge and associated diffusion from surface to subsurface and bulk in the literature. Here, we aim to understand +2 charge and neutral vacancy diffusion on anatase (101) surface using 72 and 216 atoms surface slabs employing a semi-local density functional and the Hubbard model. The activation-relaxation technique nouveau (ARTn) coupled with Quantum Espresso is used to investigate the activated mechanisms responsible for the diffusion of oxygen vacancies. The small slab model over-stabilizes the +2 charged topmost surface vacancy, which is attributed to the strong Coulomb repulsion between vacancy and neighboring Ti$^{+4}$ ions. The larger slab allows atoms to relax parallel to the surface, decreasing the +2 charged topmost surface vacancy stability. The calculated surface-to-subsurface barriers for the +2 charged vacancy and diffusion of the neutral vacancy on a slab of 216 atoms are 0.82 eV and 0.52 eV, respectively. Furthermore, the bulk vacancy prefers to migrate toward the subsurface with relatively low activation barriers 0.19 eV and 0.27 eV and the reverse process has to overcome 0.38 eV and 0.40 eV barriers for the +2 charged and the neutral vacancies. This explains the experimentally observed high concentration of vacancies at the subsurface sites rather than in the bulk, and the dynamic diffusion of vacancies from the bulk to the subsurface and from the subsurface to the bulk is highly likely on the surface of anatase (101). Finally, we provide a plausible explanation for the origin of recently observed subsurface-to-surface diffusion of oxygen vacancy from the calculated results.

## I. INTRODUCTION

Titanium dioxide, $TiO_2$, is one of the most stable, versatile, low-cost, high natural abundance, and non-toxic materials. Its rich set of properties has made it widely used as an efficient photocatalyst, pollutant degradation, antibacterial agent, hydrogen evolution, water-splitting, and $CO_2$ reduction. $TiO_2$ naturally occurs as many polymorphs, including anatase, rutile, and brookite. Among those polymorphs, rutile is the most stable form under ambient conditions. However, there is a consensus among material scientists that anatase is the most active polymorph of $TiO_2$ for photocatalysis[1]. This has led anatase to be investigated extensively, both experimentally and theoretically, using various methods, including spectroscopic methods (XPS[2], UV-vis absorption[3], APRES[4], EPR[5]) and density functional theory (DFT)[6], in order to understand the electronic structure and optical properties.

$TiO_2$ wide bandgap curtails the photocatalytic activity in the visible light and infrared regions. Hence doping, defect formation, and surface modifications are employed to improve the visible light absorption. Cation doping sites can act as recombination centres for photo-generated charge carriers, decreasing their potential to support surface redox reaction[1]. Anion doping, on the other hand, is expected to improve visible light adsorption[7]. Some researchers proposed that nitrogen (N) doping of $TiO_2$ could increase the photoactivity in the visible light region through the narrowing of the bandgap[8]. However, a significant reduction of the bandgap may lead to a material's chemical composition change that can be accompanied by structural and electronic properties change as well[9]. A recent study by Foo and co-workers found a cubic titanium oxynitride phase formation in highly nitrogen-doped anatase that provides evidence for such a chemical composition change[10]. However, the presence of oxygen vacancies is also correlated with the enhanced visible light photoactivity of doped $TiO_2$ [11]. Thus, researchers have turned their focus towards oxygen vacancies: in some cases, these produce a narrow band of defect-derived states experimentally observed at about 0.9 eV below the conduction band edge in agreement with ultraviolet photoemission spectroscopy (UPS) and electron energy loss spectroscopy (EELS) studies[12-14]. In addition, X-ray photoelectron spectroscopy (XPS) and conductivity measurements at the defective surfaces show formation of $Ti^{+3}$ ions and reduction of conductivity also evidence the localized nature of the excess electrons[2,15,16]. Among these vacancy states, the bridging oxygen vacancy formation has been largely supported by spectroscopic and chemical evidence on $TiO_2$ samples sputtered and annealed in ultra-high vacuum [17]. Moreover, comparative studies of anatase and rutile show anatase surfaces have a lower tendency to release lattice oxygen atoms whereas rutile surfaces form a large amount of oxygen defects [18,19].

Among all surfaces, anatase (101) is most widely investigated due to its superior stability and higher photocatalytic ability. Cheng and Selloni provided the first theoretical evidence for the subsurface diffusion of a neutral surface vacancy with a 0.74 eV barrier[20]. Scheiber *et al*. confirmed that theoretical prediction using electron bombardment experiment of anatase (101) surfaces[17]. Scheiber and co-workers assumed an activation energy barrier ranging from 0.6 eV to 1.2 eV for subsurface diffusion and the authors argued that such a range is not unreasonable as the vacancy diffusion barrier depends on the immediate environment of each vacancy. The metastable surface neutral oxygen vacancy formation involves dissociation of shorter bonds of Ti-O at the anatase (101) surface creating one fivefold, and one highly unstable fourfold coordinated $Ti^{+3}$ cation[20]. Conversely, removal of an oxygen atom from the subsurface creates only stable fivefold coordinated $Ti^{+3}$ cations. Thus, the subsurface vacancies are more stable than those at the (101) surface. However, Setvin *et al*. demonstrated for the first time that the relative stability of subsurface and surface vacancies on anatase (101) could be reversed by adsorbing an oxygen molecule on the surface and by applying an electric field under experimental conditions[21]. The authors elucidated the migration of the bulk oxygen vacancy to the surface, and showed that the vacancy interacts with the adsorbed oxygen species. Recent theoretical investigations of water ($H_2O$) and hydrogen sulfide ($H_2S$) adsorption on the anatase (101) surface also observe similar changes of the relative stability and demonstrate subsurface vacancy diffusion towards the surface to interact with the adsorbate species[22,23]. Moreover, electric field induced subsurface vacancy migration towards the surface suggests vacancy charge might be responsible for such stability change in addition to the surface relaxations. These studies demonstrate the complex nature of anatase (101) surface. Yet, so far, there is no consensus among researchers regarding the bulk vacancy stability with respect to the subsurface vacancy. Cheng and Selloni reported a 3.69 eV (GGA-PBE) neutral vacancy formation energy in the bulk, which is very similar to their reported value of subsurface vacancy formation energy using 216 atoms slab with six layers[19,20]. The authors also computed bulk diffusion barrier using anatase bulk unit cell and found ~0.17 eV activation barrier. However, no information was presented about bulk to subsurface vacancy migration on 216 atoms slab. Deak and co-workers calculated neutral vacancy formation energy on bulk using 72 atoms slab with three layers[24]. The reported bulk formation energy is 5.35 eV (HSE06), substantially larger than their subsurface and surface neutral vacancy formation energies. Li *et al*. also reported bulk vacancy formation energy for neutral vacancy on anatase 144 atoms slab with four layers[25]. The calculated bulk vacancy formation energies of 4.27 eV (GGA-PBE) and 4.95 eV (sX-LDA modified screened exchanged functional) were found to be larger than subsurface vacancy formation energies, but smaller than the topmost surface vacancy formation energies. Thus, our understanding of vacancy formation and diffusion on anatase (101) is far from complete.

The wide range of surface to subsurface diffusion barriers, inconsistencies of the results for bulk vacancy stability and lack of diffusion barriers associated with both +2 charged and neutral vacancies warrant this investigation on the anatase (101) surface. Despite a number of studies related to electronic structure, vacancy formation energy, and defect states in the bulk and surfaces of anatase, a few have focused on the link between the oxygen vacancy charge and its diffusion from surface to subsurface and subsurface to bulk. Therefore, in this work, we investigate the influence of vacancy charge on the stability of vacancies using neutral and +2 charged vacancies on the anatase surface, subsurface and bulk *in silico* using 72 and 216 atoms surface supercells. Semi-local density functional and the Hubbard model are employed to understand the vacancy formation energies and the activation-relaxation technique nouveau (ARTn) in its recent implementation with Quantum Espresso software suite is used to elucidate activation mechanisms responsible for +2 charged and neutral oxygen vacancy diffusion[26].

The outline of the article is as follows. First, we describe the computational details for oxygen vacancy formation energy and diffusion barrier calculations in Sec. II, then we focus on result and discussion in Sec. III with a brief comparison of semi-local functional (GGA-PBE) results with Hubbard model (DFT + *U*). Section III is consisting of two subsections where we describe the results obtain from 216 atoms large slab model for surface vacancy diffusion to subsurface in subsection A. Then we turn our attention into bulk vacancy diffusion to subsurface in subsection B. Finally, we summarize the import findings as conclusions in Sec. IV.

## II. COMPUTATIONAL DETAILS

In TiO$_2$ each Ti atom donates 4 electrons to two oxygen atoms, resulting in Ti$^{+4}$ and O$^{2-}$ nominal charges. Upon the loss of an O atom in a TiO$_2$, the electron pair originally trapped by the oxygen are donated to the lattice and remain trapped in a cavity of oxygen vacancy (V$_O$) that gives rise to a color center[1]. These electrons are commonly named excess electrons in the literature [16]. A neutral color center is equivalent to a pair of electrons associated with the V$_O$. For the sake of simplicity, a neutral color center is referred to as a neutral vacancy throughout the article; electrons left in the vacancy interact with adjacent Ti$^{+4}$ ions to give Ti$^{+3}$ centers while an electron pair deficient oxygen vacancy is referred to as a doubly charged color center[1]. In this article, a doubly charged color center is named as a +2 charged vacancy to minimize technological jargon. In addition, 72 atoms and 216 atoms slabs are referred to both non-defective and defective slabs unless otherwise specified.

Initial structural relaxation of the anatase (101) surfaces and bulk structures is performed using spin-polarized GGA-PBE exchange-correlation functional within plane-wave-pseudopotential scheme implemented in Quantum Espresso (QE) software suite [27,28]. The neutral and the +2 charged states of the oxygen vacancies (denoted V$_O^0$ and V$_O^{+2}$) are used as the intrinsic defects. A model structure with three O-Ti-O layers is used to simulate the TiO$_2$ (101) stoichiometric surface with (1×2) supercell of 72 atoms (Ti$_{24}$O$_{48}$). The supercells are further tested for a 2×2×1 k-point mesh but the difference in formation energies of defects is found to be negligible with respect to gamma point calculations. Thus, all the vacancy formation energies reported here are obtained from gamma point only calculations. A vacuum layer of 12 Å is used to separate the periodic images along the z-axis direction for all the slab models. The interaction between core and valence electrons is described by ultrasoft pseudopotentials and a kinetic energy cutoff of 50 Ry for smooth part of the wave functions is used[29]. $2s^2sp^4$ and $3s^23p^63d^24s^2$ are treated as the valence electrons of oxygen and titanium, respectively. The computed bulk lattice parameters a = 3.78 Å and c = 9.45 Å (experimental values: a = 3.77 Å and c = 9.50 Å [30]) are obtained by relaxing a tetragonal bulk unit cell. A 5×5×2 k-points mesh is used to sample the corresponding Brillouin zone for the four $TiO_2$ units containing bulk unit cell. In addition to the 72 atoms surface supercells, 216 atoms ($Ti_{72}O_{144}$) four O-Ti-O layers with (2×3) large surface supercells are employed for anatase (101) surface studies. A bulk oxygen vacancy is also computed using a 3× 3 ×2 supercell with 216 atoms. As with the 72-atom system, k-point sampling is limited to the gamma point for 216-atom surface slabs and bulk supercells. We perform test calculations on 72 atom anatase (101) surface supercells and find that the two unpaired electrons associated with the $Ti^{+3}$ in neutral oxygen vacancy can be represented as singlet and triplet states that are degenerate at the generalized gradient approximation (GGA) level using Perdew-Burke-Ernzerhof (PBE) exchange-correlation functional[31]. However, standard DFT functionals tend to delocalize electrons over the crystal particularly for the material that contain transition elements with partially filled *d* or *f* electronic orbitals, thus, Hubbard-like correction term (*U*) is applied to *d* orbitals of Ti in anatase[32,33]. The DFT+*U* calculations are performed on both 72 and 216 A-$TiO_2$(101) slabs using gamma point with the PBE functional and spin-polarized scheme. After extensive testing of the Hubbard *U* parameter using 72 atoms slab, U = 2 eV is selected for the optimizations and diffusion barriers calculations due to destabilization of subsurface vacancy beyond this value for the neutral oxygen vacancies on the anatase (101) surface. This is graphically illustrated in FIG. S1 of the supplementary materials section. During the structural optimizations, all the atoms of the slab are allowed to move except those in the bottom layer. The Broyden-Fletcher-Goldfarb-Shanno algorithm is employed for all the ionic steps during the optimization with the fixed occupations due to semiconducting nature of this material[34]. For density of states (DOS) and projected density of states (PDOS) calculations 8×8×1 larger k-point mesh is used for both DFT and DFT+$U$. Procedure documented in the Quantum Espresso user guide is followed starting from the pre-optimized defective 72 atoms anatase (101) surface structure for the computation of PDOS and DOS.

Neutral oxygen vacancies ($V_O$) in anatase (101) surface are simulated by removing one oxygen atom from both 72 and 216 atoms supercells. A charged vacancy ($V_O^{+2}$) is created by removing one oxygen atom and two electrons from the supercells (homogeneous background charge is added to compensate for extra charges to remove electrostatic divergence if the periodic cell is not neutral)[35]. To determine the relative stability of neutral and +2 charged states, the formation energies ($E_{form}$) of $V_O$ and $V_O^{+2}$ defects are computed assuming thermodynamic equilibrium with a reservoir of oxygen (oxygen chemical potential $\mu_O$). The oxygen chemical potential is conveniently referred to the energy $E_{tot}$ of an O atom in $O_2$, which is equivalent to $1/2 E_{tot}(O_2) + \mu'_O$. In this work, $\mu'_O = 0$ is assumed, which corresponds to the O-rich limit at which oxygen condensation occurs. Therefore, formation energy is defined as follows: [15]

$$E_{form}(V_O^q) = E_{def}^q - E_{no-def} + \tfrac{1}{2} E_{tot}(O_2) + \mu'_O + q(E_V + E_F), \qquad (1)$$

where q is charge state of the oxygen vacancy, $E_{def}^q$ and $E_{no-def}$ are total DFT/DFT+$U$ energy of the defect containing supercell with vacancy charge q and pristine supercell without the defect, respectively. $E_F$ and $E_V$ are the Fermi energy of the material which is the chemical potential of the electron reservoir and the energy of the valence band maximum, respectively. Madelung corrections for the electrostatic interaction between homogeneous background charge and the charge defect are not calculated in the present work since the dielectric constant of the $TiO_2$, in particular, the ionic contributions are fairly large[15].

All structures are optimized before proceeding with diffusion barrier calculations. Diffusion pathways and barriers are calculated for both the GGA-PBE and the DFT+$U$ levels for 72-atom supercells. However, GGA-PBE and DFT+$U$ are tested and used only for a few activation barrier computations on 216-atom slabs. Here, the activation-relaxation technique nouveau (ARTn) coupled to the QE code is utilized to compute the activation energies and diffusion pathways [26,36-38]. Unlike string methods such as nudged elastic band (NEB), that require saddle point and final state configurations of the material as initial guess structures in addition to the initial state, ARTn requires only initial minimum energy configuration of the material to search the minimum-energy pathway (MEP) and associated activation barriers. Therefore, prior knowledge of the complex reaction pathways and of their structural configurations is not needed to apply ARTn method. In the novel implementation of ARTn, which is referred to as ab-initio-ARTn, the DFT based energy and force engine of the QE is used for energy and force evaluation [26]. In this work, fully relaxed initial configuration of the vacancy structure is used to initiate the MEP search and the user selects the direction of the oxygen atom along which the atom and its neighbouring environment is pushed iteratively based on the vacancy location to reduce the computational time. ARTn events are generated as follows: the target oxygen atom is moved in a user defined direction iteratively. At each step, a minimum relaxation is performed in the perpendicular hyperplane to prevent collision and allow the whole system to react to this displacement. Before a new step, the local curvature is computed using a Lanczos algorithm in order to extract the lowest eigenvalue and its corresponding eigenvector and this step is repeated, with a small push in the defined direction, as long as the lowest eigenvalue is above a given threshold (set here to 0.03 eV/Å)[39]. Once the system leaves the initial harmonic basin, as signaled with a negative eigenvalue, the system is then pushed along the associated direction of negative curvature, while minimizing in the perpendicular direction. No constraint is placed, at this point, on the moving atoms which are iteratively pushed in the user defined direction, while a minimal relaxation is performed in the perpendicular hyperplane, until being out of its harmonic basin. This movement induces local deformation around the selected oxygen atom and its nearest neighbors. Then, the lowest eigenvalue of the Hessian matrix is calculated using Lanczos algorithm[39]. This continues until the lowest eigenvalue falls below a threshold value, which indicates the system has left the harmonic basin and moved beyond the inflection line. Then the activation part of ARTn begins to push from the inflection to the saddle point following the direction defined by the eigenvector corresponding to the lowest eigenvalue. After each push along the oxygen atom movement direction, the whole system is relaxed in the hyperplane perpendicular to the above direction. This iterative procedure, including the calculation of the curvature, is repeated until the total force falls below a given threshold defining the saddle point. Here, 0.03 eV/Å is used as a threshold value for total force to define convergence at the saddle point. From the saddle point, the configuration is further pushed along the eigenvector both away from the initial minimum and towards it. The two configurations obtained from the two pushes are then fully relaxed and compared with the initial minimum. The latter step ensures that the saddle point is directly connected to the initial minimum without an intermediate state(s). Thus, it confirms the fully connected event from initial state to saddle point and to final state. Interested readers are directed to the reference 26 for details of the method.

## III. RESULTS AND DISCUSSION

Comparing GGA-PBE optimized 216-atom and 72-atom anatase (101) surface structures with the recently elucidated experimental bond lengths of the anatase surface (see FIG. S2 for atom numbering scheme and TABLE S1 in the supplementary materials section), we find that our theoretically predicted surface structures match well with the experimental bond lengths [40]. The maximum difference between the experimental and the theoretically predicted bond lengths of the 72-atom slab surface is 0.16 Å and that is of the 216-atom slab is 0.23 Å; overall, the theoretical structures display slightly elongated bonds. We attribute the 0.23 Å difference in the larger four-layer slab model to stronger structural relaxation compared to the three-layer 72-atom slab. A side view of the 72-atom and 216-atom supercells with the vacancy numbering scheme is depicted in FIG. 1, and the vacancy formation energies of the neutral vacancies computed from the GGA-PBE level of theory are listed in TABLE I. The vacancy numbering scheme in FIG. 1 shows the vacancy numbers as a subscript depending their relative height as measured from the surface while the vacancy charge is represented as a superscript. For example, topmost neutral surface oxygen vacancy is denoted as $V^0_{O1}$, and the deepest subsurface neutral oxygen vacancy in the 72-atom slab of the anatase (101) surface is denoted as $V^0_{O5}$. The small slab model is computationally less demanding and widely used in literature [19,20,24,41]. Thus, the 72-atom slab is initially used to explore formation energies and activation energy barriers of oxygen vacancies on anatase (101) surface with inequivalent oxygen vacancy sites.

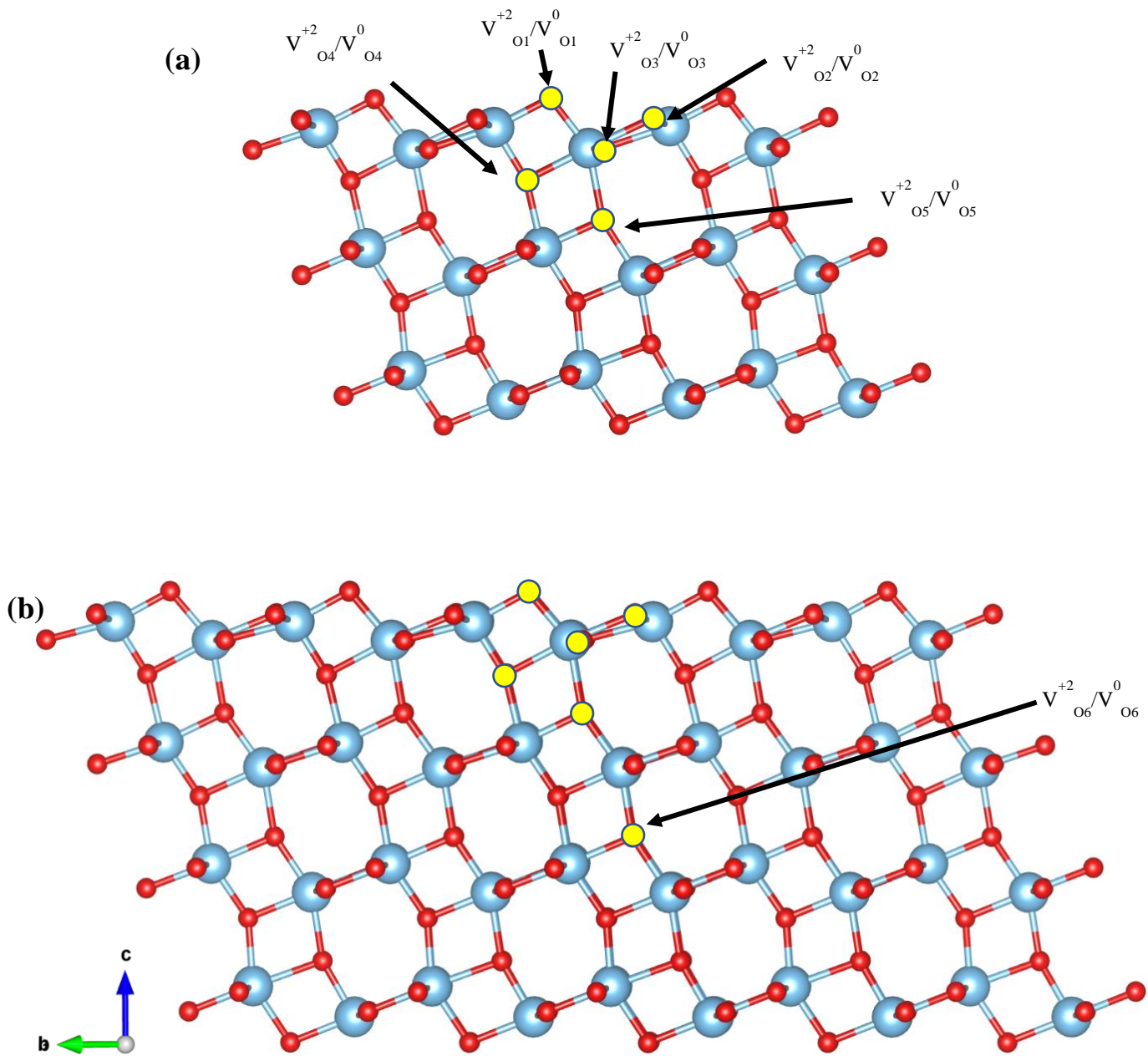

**FIG. 1.** Side view of the anatase (101) surface supercell of 72 atoms (a) and 216 atoms (b). Oxygen vacancy positions are labelled for easy references. The larger 216-atom slab is used to compute bulk vacancy formation energy as well, which is denoted as $V^0_{O6}$ for neutral vacancy and $V^{+2}_{O6}$ for +2 charged vacancy. Vacancies are presented as yellow dots, oxygen red and titanium blue.

TABLE I. Formation energies $E_{form}$ of neutral defects at different surface and subsurface sites for an anatase (101) slab of 72 atoms and 216 atoms calculated at the GGA-PBE level for oxygen rich conditions. Our results are compared with the literature reported values. Here, $V^0_{O6}$ represents the bulk neutral oxygen vacancy formation energy computed from the 216-atom slab model.

| Defect site | 72 atoms $E_{form}$(eV) | 72 atoms $E_{form}$(eV) | 96 atoms $E_{form}$(eV) | 144 atoms $E_{form}$(eV) | 144 atoms $E_{form}$(eV) | 216 atoms $E_{form}$(eV) | 216 atoms $E_{form}$(eV) |
|---|---|---|---|---|---|---|---|
| $V^0_{O_1}$ | 4.24 | 4.39 | - | 4.18 | 5.07 | 4.01 | 4.15 |
| $V^0_{O_2}$ | 5.51 | 5.43 | - | 4.89 | 5.56 | 5.00 | - |
| $V^0_{O_3}$ | 5.07 | 4.95 | - | 4.54 | 5.35 | 4.37 | - |
| $V^0_{O_4}$ | 4.61 | - | - | - | - | - | - |
| $V^0_{O_5}$ | 4.04 | 4.18 | 5.12 | 4.00 | 4.91 | 3.71 | 3.69 |
| $V^0_{O_6}$ | - | - | - | - | - | 3.83 | - |
| Bulk | - | - | - | - | - | 3.79 | 3.69 |
|  | This work | Ref. [20] | Ref. [42] | Ref. [22] | Ref. [23] | This work | Ref. [20] |

The vacancy formation energy of the subsurface layer vacancy $V^0_{O5}$ is the lowest among all the considered neutral vacancies on the anatase (101) surface. In the 72-atom slab, the topmost surface vacancy $V^0_{O1}$ is ~0.2 eV higher in energy than $V^0_{O5}$, quantitatively consistent with the previous theoretical studies of 72-atom and 144-atom slabs [19,20,22,23]. The $V^0_{O4}$ vacancy is considered very unstable in previous studies; however, we find that this vacancy is metastable as it has to overcome activation energy barrier for diffusion. The 216-atom slab shows ~0.3 eV energy difference between the topmost surface and the subsurface oxygen vacancy formation energy. The bulk neutral vacancy stability $V^0_{O6}$ stands between the topmost surface vacancy and the subsurface vacancy. The bulk neutral vacancy formation energy is computed using 216-atom bulk supercell and agrees well with its evaluation in the 216-atom surface slab. Deak *et al*. recently computed the vacancy formation energy for the topmost layer, the subsurface and the bulk of 72 atoms supercell with HSE06 hybrid functional and found that the bulk vacancy formation energy is the highest among all[24]. Li *et al*. also reported surface, subsurface and bulk vacancy formation energies from GGA and screened hybrid functional [25]. The authors show that the bulk vacancy formation energy is higher than the subsurface, but lower than the topmost surface from both methods. Thus, our bulk vacancy formation energy is also in agreement with Li and co-workers. TABLE S2 shows the vacancy formation energies for +2 charged vacancies, which are given at the higher Fermi energy level on the conduction band minimum (CBM) in oxygen-rich conditions. Note that the $V^{+2}_{O5}$ subsurface vacancy is 0.06 eV higher in energy than the $V^{+2}_{O1}$. Therefore, the +2 charged topmost layer vacancy on the anatase (101) surface is slightly more stable than the subsurface $V^{+2}_{O5}$ on the 72- atom slab. In addition, the $V^{+2}_{O4}$ vacancy is not observed as the $V^{+2}_{O4}$ vacancy diffuses to the $V^{+2}_{O1}$ vacancy site during the structural optimization, along a barrier-less trajectory. On the contrary, a larger 216-atom slab shows a subsurface vacancy 0.19 eV more stable than that of the topmost surface vacancy. Computed bulk vacancy formation energy of the $V^{+2}_{O6}$ is quantitatively identical to the topmost surface vacancy formation energy. The bulk +2 charged vacancy formation energy computed from 216-atom bulk supercell is also comparable with the $V^{+2}_{O6}$ bulk vacancy formation energy. Generally, $V^0_{O2}$, $V^0_{O3}$, $V^{+2}_{O2}$, and $V^{+2}_{O3}$ vacancies have significantly higher vacancy formation energies than $V^0_{O1}$, $V^0_{5}$, $V^{+2}_{O1}$ and $V^{+2}_{O5}$.

Semi-local functionals such as PBE tend to delocalize electrons [6]. Therefore, the DFT+$U$ method is applied to all the above vacancies in both neutral and +2 charged states and is tested for the most suitable Hubbard parameter value using 72 atoms slab. Starting from the GGA-PBE optimized anatase (101) surface structures, a range of Hubbard values from 1 eV to 7 eV is applied. For neutral vacancies, above 2 eV and higher Hubbard parameter stabilizes the surface $V^0_{O1}$ vacancy; thus, stability crossover from subsurface-to-surface oxygen vacancy is observed. Nevertheless, according to the theoretical, and experimental observations, neutral subsurface oxygen vacancy $V^0_{O5}$ should be more stable, consequently, the top-layer surface oxygen vacancy should diffuse to the subsurface[17,20]. As a result, 2-eV Hubbard parameter value is chosen for the DFT + $U$ calculations. The applied Hubbard values do not show such a stability crossover effect in the +2 charged vacancies. This can be attributed to the higher stability of $V^{+2}_{O1}$ on the topmost surface layer of the 72 atoms supercell, as illustrated in FIG. S1. The applied Hubbard value-induced crossover in relative stabilities of $V^0_{O5}$ to $V^0_{O1}$ is observed by Cheng *et al*. who predicted that U = 3 eV is the best value for the anatase (101) surface studies [20]. However, we find that a 3-eV on-site Hubbard value affects profoundly the stability of the topmost surface neutral vacancies.

The projected density of states (PDOS) is calculated from the DFT +$U$ (U = 2 eV). Although this Hubbard value underestimates the bandgap of the anatase (101) surfaces consisting of the neutral and the +2 charged oxygen vacancies, electron localization at the mid-gap states are observed for the neutral charge oxygen vacancy (see FIG. S3 in supplementary materials). Therefore, a 2 eV Hubbard parameter is still sufficient to localize electrons at vacancy site on the anatase (101) surface. The localization of electrons occurs at 0.48 eV and 0.46 eV below CBM for the $V^0_{O5}$ vacancy. These values are comparable with the literature [20]. Although the 2-eV Hubbard parameter satisfies the electron localization, the defect states in the band gap occur closer to the CBM. Applying higher Hubbard values such as 3 eV and 3.5 eV move the gap states deeper into the band gap 0.55 eV, 0.59 eV and 0.86 eV, 0.91 eV, respectively. However, we cannot use such high U values of the Hubbard parameter as they significantly affect the stability order of the vacancies (FIG. S1). The selected 2-eV Hubbard U value does not alter the defect formation energies considerably since the anatase valence band I completely filled. Therefore, these excited electronic states at the band gap cannot fallback into the valence band. Moreover, electrons are in localized states sufficiently below the CBM so that it is unlikely that they will diffuse into the conduction band as well. FIG. S4 provides vacancy formation energies for all the +2 charged and neutral vacancies in 72-atoms (FIG. S4 (a)) and 216-atoms supercells (FIG. S4 (b)) of the anatase (101) surfaces computed with DFT + $U$ at a U value of 2 eV. As can be seen, +2 charged vacancy formation is favorable at all the Fermi levels. The FIG. S4 (a) clearly shows higher stability of $V^{+2}_{O1}$ vacancy site on the smaller 72 atoms slab. Therefore, both GGA-PBE and DFT + $U$ methods predict higher stability on the top-layer surface +2 charged vacancy on the 72-atom slab. This implies the 72-atom slab is not adequate to represent +2 charged anatase (101) surface vacancy. The larger 216-atom slab allows to relax atoms parallel to the surface which reduces the +2 charged topmost surface vacancy stability and leads to $V^{+2}_{O5}$ subsurface vacancy becoming the most stable vacancy site. Since all the other vacancy formation energies and diffusion barriers computed on smaller slab agree well with the larger slab model, the 216-atom is primarily used for the discussion. We note, also that the bulk vacancy formation energies computed on a 216-atom bulk supercell and a 216-atom surface slabs ($V^{+2}_{O6}$/ $V^{0}_{O6}$) are higher than the topmost surface and the subsurface vacancy formation energies. Thus, the DFT + $U$ computed bulk vacancy formation energies deviate from the GGA-PBE results. However, a hybrid functional study with HSE06 shows similar trend comparable with our DFT + $U$ result[24]. This suggests the vacancy formation energy is very sensitive to the computational method and functionals.

Although extra charge distribution in 3D structures is very well compensated by the "Jellium" background charge to keep the total charge of the system neutral, spurious charge states interacting with the Jellium can cause an error in the absolute value of defect formation energy for charged defects, in particular, low dimensional systems and supercell slab models with a vacuum [43,44]. Such shortcomings can affect the validity of the comparison of the +2 charged vacancy formation energy with the neutral vacancy formation energy. Comparing relative vacancy formation energies with respect to the subsurface vacancy formation energy based on the total energy of each defect structure, we find relative stability order and the magnitude of the energy difference between each vacancy for both vacancy types. The errors in the vacancy formation energies due to the compensating background charge interaction with the charged defect cancel out in such an analysis (not shown). Moreover, a recent study concluded the anatase (101) surface charge vacancy formation energy depends on the choice of the vacuum spacing as well [45]. As we do not apply a correction scheme, the 12 Å vacuum spacing used in this work may overestimate the 2+ charged vacancy formation energy by 0.2 eV. However, the 2+ charge state is more favorable at all Fermi levels is not sensitive to an error of this size. Furthermore, our vacancy formation energies agree well with the previous studies [20,25]. Therefore, our charged vacancy formation energies can be compared with the neutral vacancy formation energies.

**TABLE II**. Diffusion barriers ($E^A$) of neutral and +2 charged defects on the different surface and subsurface sites of an anatase (101) slab of 72 atoms calculated at the GGA-PBE level using ARTn coupled to QE.

| Pathway | $E^A$ neutral vacancy direct pathway (eV) | $E^A$ neutral vacancy inverse pathway (eV) | $E^A$ +2 charged vacancy direct pathway (eV) | $E^A$ +2 charged vacancy inverse pathway (eV) | Ref. 20 (eV) direct pathway | Ref. 20 (eV) Inverse pathway |
|---|---|---|---|---|---|---|
| $V_{O1} \rightarrow V_{O5}$ | 0.60 | 0.82 | 0.76 | 0.72 | 0.74 | 0.95 |
| $V_{O1} \rightarrow V_{O3}$ | 1.44 | 0.59 | 1.42 | 0.31 | 1.34 | 0.78 |
| $V_{O1} \rightarrow V_{O2}$ | 3.00 | 1.73 | 3.45 | 1.70 | 1.52 | 0.47 |
| $V_{O2} \rightarrow V_{O3}$ | 0.11 | 0.54 | 0.03 | 0.66 | 0.30 | 0.78 |
| $V_{O3} \rightarrow V_{O5}$ | 0.71 | 1.76 | 0.64 | 1.70 | 0.78 | 1.59 |

**TABLE III**. Diffusion barriers ($E^A$) of neutral and +2 charged defects on selected surface and subsurface sites of an anatase (101) slab of 216 atoms calculated at the GGA-PBE level using ARTn coupled to QE.

| Pathway | $E^A$ neutral vacancy direct pathway (eV) | $E^A$ neutral vacancy inverse pathway (eV) | $E^A$ +2 charged vacancy direct pathway (eV) | $E^A$ +2 charged vacancy inverse pathway (eV) |
|---|---|---|---|---|
| $V_{O1} \rightarrow V_{O5}$ | 0.52 | 0.81 | 0.82 | 1.01 |
| $V_{O1} \rightarrow V_{O3}$ | 1.14 | 0.78 | 1.15 | 0.73 |
| $V_{O3} \rightarrow V_{O5}$ | 0.67 | 1.33 | 0.73 | 1.34 |
| $V_{O5} \rightarrow V_{O6}$ | 0.40 | 0.27 | 0.38 | 0.19 |

**A. Surface and subsurface diffusion pathways**

The diffusion barriers computed with the GGA-PBE method are presented in TABLE II for 72 atoms smaller slab and in TABLE III for 216 atoms larger slab. In order to keep the calculations consistent, mechanisms are described from the higher energy vacancy position to the lower energy vacancy position migration throughout the article, except for the surface to subsurface diffusion on 216 atoms slab. All the GGA-PBE level calculated potential energy profiles and mechanisms for the 72 atoms supercells with single point defect are illustrated from FIG. S5 to S9 in supplementary materials sections. Corresponding calculations from the DFT + $U$ level are illustrated in FIG. S10 (a) to (h), and activation energy barriers are also tabulated in TABLE S3 and S4 for the small and the large supercells, respectively. Remarkable similarity in the potential energy profiles and the vacancy migration barriers computed from semi-local functional and Hubbard model suggests that, in spite of its aforementioned limitations, GGA-PBE level computations are sufficient to accurately predict diffusion barriers and corresponding vacancy migration mechanisms. However, we note that $V^0_{O1}$ to $V^0_{O5}$ surface-to-subsurface diffusion barriers computed from the DFT + $U$ method for both smaller and larger slab models are significantly smaller than the GGA-PBE level computed values. That may be attributed to the strong localization of two electrons (polaron formation) at the neighboring $Ti^{+3}$ ions surrounding the topmost surface vacancy site[24]. Structural deformation associated with the electron localization at the vacancy site may lower the activation barrier. Nevertheless, this is beyond the scope of the present work as our primary objective is to establish the vacancy diffusion directions on the anatase (101) surface and the barriers associated with the +2 charged and the neutral vacancies.

Here, first we focus on surface-to-subsurface diffusion barriers and pathways on +2 charged and neutral oxygen vacancies. The neutral and the +2 charged vacancies diffusion from the surface to the subsurface show the expected behavior on 216-atom slabs. The neutral vacancy diffusion barrier 0.52 eV is relatively lower than that of the +2 charged vacancy 0.82 eV for the large supercell, respectively. FIG. 2 illustrates subsurface to surface diffusion pathway and potential energy profile associated with 216 atoms slab calculation. Here, the subsurface most stable vacancy site migration to the topmost surface is computed in order to compare with 72-atom slab neutral vacancy calculation. All the computed subsurface to topmost surface diffusion involves two steps; first the subsurface (or surface) vacancy migrates to the $V^{+2}_{O4}/V^{0}_{O4}$ site, then the vacancy escapes from this unstable state and diffuses to the final surface (subsurface) vacancy site. During this process two oxygen atoms are displaced substantially, which are colored in black for easy visualization and the vacancy site represents as a yellow circle. The potential energy profiles represent the relative stability of the vacancy states and the activation barrier. In addition to this mechanism, $V^{+2}_{O3}/V^{0}_{O3}$ surface vacancies can migrate to $V^{+2}_{O5}/V^{0}_{O5}$ sites. Larger 216 atoms slab calculation reveals activation energy barriers 0.73 eV and 0.67 eV for the +2 charged and the neutral vacancies, respectively. These calculated values for surface-to-subsurface vacancy diffusions are consistent with the theoretical result and experimental range of 0.6 eV to 1.2 eV further convince GGA-PBE level of theory is sufficient for the activation barrier calculations[17,20].

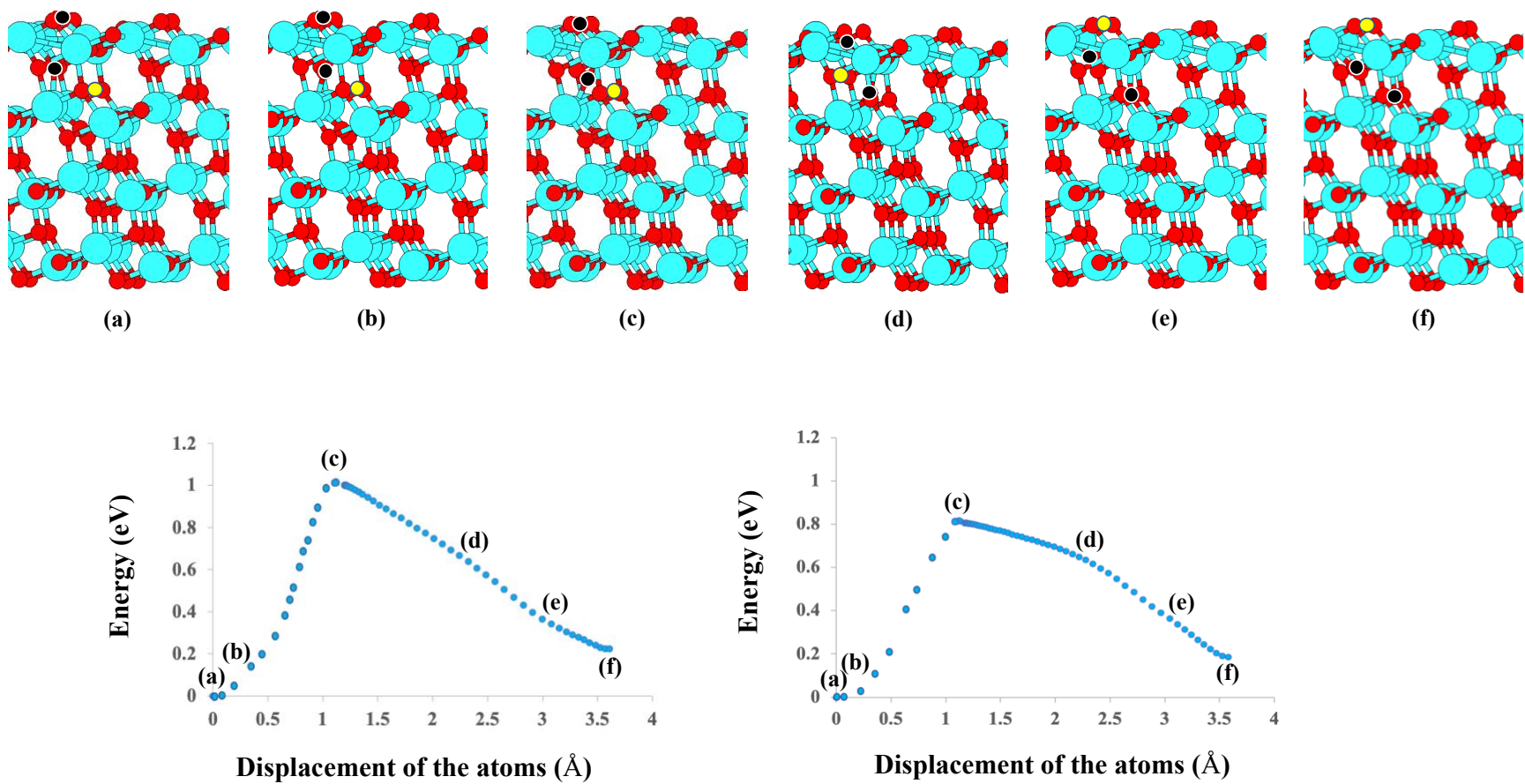

**FIG. 2.** Potential energy profile along $V^{+2}_{O5}/V^{0}_{O5} \rightarrow V^{+2}_{O1}/V^{0}_{O1}$ computed at GGA-PBE level of theory with coupled ARTn/QE on 216 atoms slab. Upper panel shows diffusion pathway with selected atomic configurations. Both +2 charged and neutral vacancy pathways are similar and only one is illustrated here. Diffusion begins with a single bond dissociation of $O_{3c}$ atom, creating an under-coordinated surface Ti ion ($Ti_{4c}$). Displacement of $Ti_{4c}$ ion vertically opens up the passage to downward migration of the surface bridging oxygen. This oxygen eventually forms sub-bridging oxygen immediately below the topmost surface vacancy. The vacancy migrates through an intermediate unstable state $V_{O4}$; hence a characteristic two-step diffusion mechanism is observed. Potential energy profile of the +2

charged vacancy is on the left, and that is for neutral vacancy on the right. Two oxygen atoms that undergo large displacement are black, Ti light blue, the oxygen red and oxygen vacancy yellow.

In order to completely describe vacancy migration processes on the anatase (101) surface, we also compute surface vacancy diffusion. The larger 216-atom slab calculation confirms the metastable nature of $V^{+2}_{O2}$ / $V^{0}_{O2}$ vacancies, which is observed for smaller slab model. In fact, DFT + $U$ level calculations show barrier-less vacancy diffusion from $V^{+2}_{O2}$ towards $V^{+2}_{O3}$; hence we do not take $V^{+2}_{O2}$/$V^{0}_{O2}$ vacancies into account for activation energy barrier calculations on the larger slab models. Instead, we focus on $V^{+2}_{O3}$/$V^{0}_{O3}$ vacancies migration to $V^{+2}_{O1}$/$V^{0}_{O1}$. The energy barrier for +2 charged $V^{+2}_{O3}$ vacancy site diffusion towards $V^{+2}_{O1}$ is 0.73 eV and for the similar diffusion pathway for the corresponding neutral vacancy is 0.78 eV. Overall, anatase (101) surface neutral vacancy migration occurs from $V^{0}_{O3}$ to $V^{0}_{O1}$ as the first step, and then $V^{0}_{O1}$ metastable vacancy diffuses to $V^{0}_{O5}$ subsurface site. $V^{0}_{O3}$ to $V^{0}_{O5}$ pathway is also possible due to lower activation barrier than $V^{0}_{O3}$ to $V^{0}_{O1}$ pathway. Surface vacancy $V^{+2}_{O3}$ with +2 charged state has equal probability to diffuse either to metastable $V^{+2}_{O1}$ site or directly diffuses to $V^{+2}_{O5}$ as these competing migration paths have identical 0.73 eV activation energy barrier. Therefore, a low barrier path of $V^{+2}_{O3}$ to $V^{+2}_{O5}$ is favorable for +2 charged vacancy surface-to-subsurface diffusion. However, $V^{+2}_{O1}$ also diffuses to the subsurface overcoming 0.82 eV barrier so that both surface vacancy types eventually migrate to the subsurface. Considering more accurate diffusion directions obtained from 216 atoms slab calculation, we conclude that the experimental diffusion range is a result of both different charges on the vacancies and competing mechanisms on the anatase (101) surface.

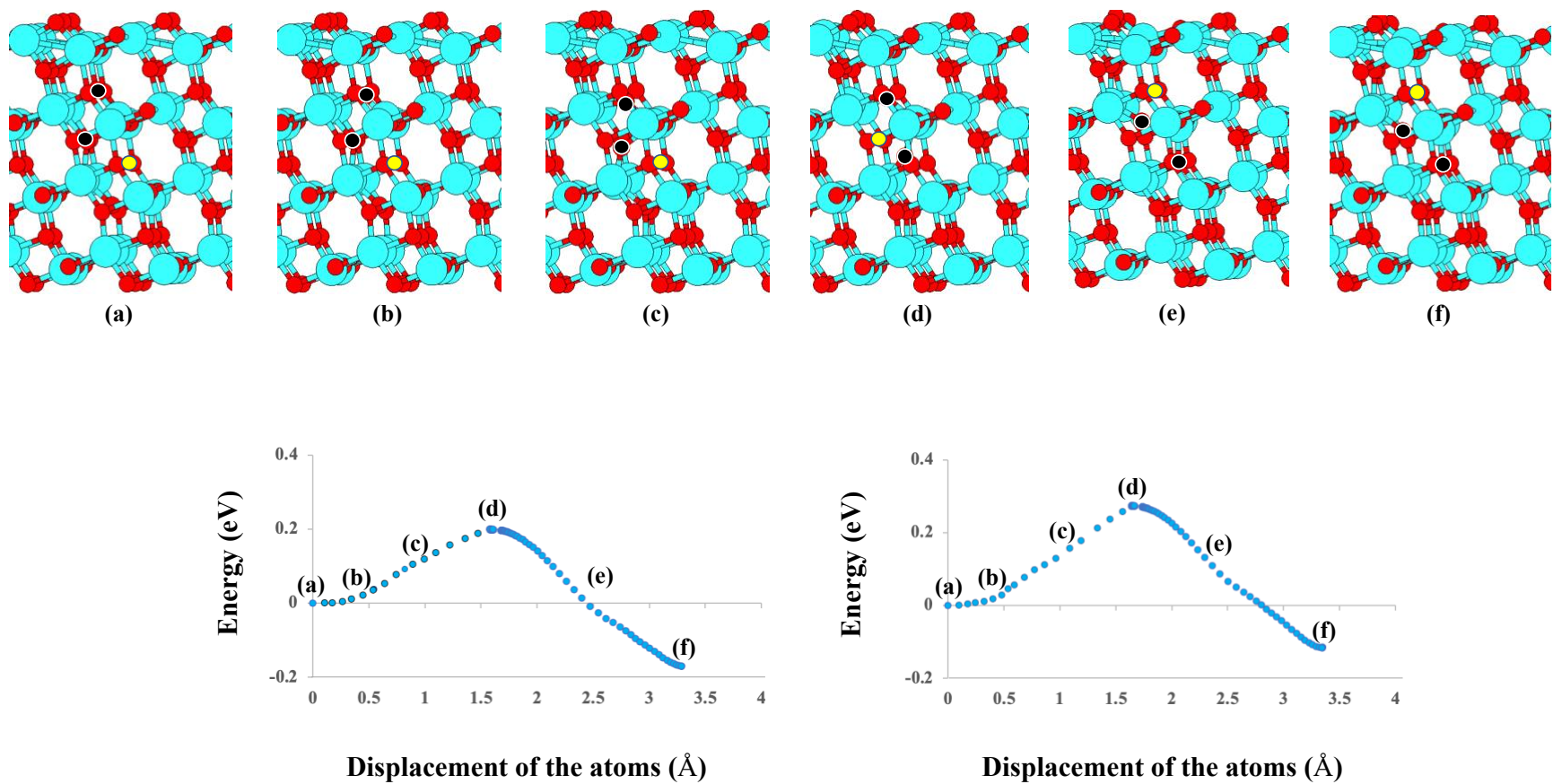

**FIG. 3.** Potential energy profile along $V^{+2}_{O6}/V^{0}_{O6} \rightarrow V^{+2}_{O5}/V^{0}_{O5}$ computed at GGA-PBE level of theory with coupled ARTn/QE on 216 atoms slab. Upper panel of the figure shows diffusion pathway with selected atomic configurations. Both +2 charged and neutral vacancy pathways are similar and only one is illustrated here. Surrounding atoms are removed for better visualization of the diffusion pathway. Diffusion begins with a single bond dissociation of $O_{3c}$ atoms indicated in black color. These oxygen atoms significantly displaced during the vacancy migration. The vacancy migrates through an intermediate unstable state; hence characteristic two step diffusion mechanism is observed. Potential energy profile of the +2 charged vacancy is on the left, and that is for neutral vacancy on the right. The two oxygen atoms that undergo large displacement are black, Ti light blue, the oxygen red and oxygen vacancy yellow.

**B. Bulk and subsurface diffusion pathways**

With a clear understanding of surface to subsurface diffusion, we now turn to diffusion pathways between bulk and subsurface layer. The diffusion barrier of bulk vacancy migration toward (101) surface calculated using 216 atoms surface slab is largely missing in the literature. Our computed bulk vacancy formation energy on the 216 atoms slab for the neutral vacancy is larger than the subsurface vacancy formation energy, but smaller than that of the topmost surface vacancy. This is in agreement with Li and co-workers reported stability order of vacancy formation energies on anatase (101) surface slab [25]. The +2 charged bulk vacancy, however, is quantitatively similar in value to the topmost surface vacancy formation energy. The Hubbard model provides higher vacancy formation energies for the bulk than for the subsurface and the topmost surface for neutral and +2 charged vacancies (see FIG. S4 (b) in supplementary martials section). Therefore, these computed bulk vacancy formation energies indicate bulk to subsurface vacancy diffusion in anatase. The ab-initio-ARTn computed diffusion barriers for the neutral and the +2 charged vacancies are tabulated in TABLE III and potential energy profiles with the diffusion pathway are illustrated in FIG. 3. The characteristic two step mechanism through unstable intermediate vacancy site can be observed for bulk to subsurface vacancy migration as well. Activation energy barrier for the +2 charged vacancy migration from the bulk to the subsurface is 0.19 eV, whereas neutral vacancy migration has to overcome 0.27 eV barrier. Therefore, we unambiguously provide evidence for bulk to subsurface diffusion. It is interesting to see that the inverse activation energy barriers for subsurface to bulk oxygen vacancy migration are comparatively low as well. This implies that, at experimental annealing temperatures, reverse diffusion is also possible. Consequently, dynamic oxygen vacancy migration from bulk-to-subsurface and subsurface-to-bulk can be observed. Our results provide theoretical evidence for surface-to-subsurface and bulk-to-subsurface oxygen vacancy diffusion, hence experimentally observed accumulation of oxygen vacancies at the subsurface is well substantiated[46]. As can be seen, ubiquitous diffusion direction in anatase is bulk to subsurface vacancy migration. This is expected as a significant amount of energy is required to create a point defect in highly crystalline nature of the bulk anatase. Therefore, a single vacancy created in the bulk diffuses towards the (101) surface direction to lower the energy of the structure.

However, surface to subsurface vacancy diffusion is not quite intuitive. Thus, we tabulate Ti-Ti and Ti-O bond distances of the defect free and defective anatase (101) surfaces with single neutral vacancy on 216 atoms slab in TABLE IV. The topmost surface $Ti_1$-$Ti_2$ distance, $Ti_1$-$O_1$ and $Ti_2$-$O_1$ bond lengths are shorter than the second ($Ti_3$-$Ti_4$, $Ti_3$-$O_5$ and $Ti_4$-$O_5$) and third layers ($Ti_5$-$Ti_6$, $Ti_5$-$O_8$ and $Ti_6$-$O_8$) in defect free structure. Also, the topmost surface maintains shorter bond lengths and distance between Ti ions in the presence of the subsurface $V^0_{O5}$ and the bulk $V^0_{O6}$ vacancies. Moreover, the $Ti_1$-$Ti_2$ distance expands ~0.6 Å in order to accommodate $V^0_{O1}$ vacancy, which is the largest among three vacancies compared in the TABLE IV. These observations indicate the anatase (101) surface topmost layer is compressively strain. Thus, to accommodate topmost surface vacancy requires substantial amount of energy to distort the surface structure. As a consequence, topmost surface vacancy is metastable and diffuses to the subsurface. Although it is difficult to find such a distinctive structural trend to support bulk to subsurface diffusion, substantially similar bond distances ($Ti_3$-$O_5$ and $Ti_4$-$O_5$) in the defect free structure suggest the second layer is minimally strain on the anatase (101) surface. This result in a lower vacancy formation energy on the subsurface than the bulk. An oxygen atom removal from the subsurface and the bulk create stable fivefold Ti ions in both cases do not explain extra stability on subsurface, however, strain on different surface layers provide more consistent explanation. Furthermore, higher stability of the +2 charged topmost surface on 72 atoms surface slab provide plausibility to stabilize topmost surface due to vertical displacement of $Ti^{+4}_{4c}$ ion neighboring the vacancy. In fact, single titanium ion displaced towards the vacuum at the adsorption site can be observed in recent computational studies of $H_2O$ and $H_2S$, which are claimed to induce subsurface-to-surface vacancy diffusions[22,23]. This also suggests charge vacancy site on the anatase (101) surface may be reconstructed in such way, the surface can engineer to reverse the natural diffusion direction. Therefore, +2 charged vacancy diffusion is equally important to understand surface reactivity and to design novel surface engineered structures of $TiO_2$-based photocatalysts.

**Table IV.** GGA-PBE level computed Ti-Ti ions distance and Ti-O bonds on the 216 atoms non-defective slab and defective 215 atoms surface slabs with single neutral vacancy are tabulated. The neutral vacancy is used as representative case as both +2 charged and the neutral vacancies show similar diffusion directions on the 216 atoms slab. The numbering scheme of the atoms can be found in Fig. S2 in supporting materials section. $O_1$, $O_5$, and $O_8$ oxygen atoms are deleted to create $V^0_{O1}$ $V^0_{O5}$ and $V^0_{O6}$ vacancies, respectively.

| Bonds/Distance | Non-defective 216 atoms | $V^0_{O1}$ defective 215 atoms | $V^0_{O5}$ defective 215 atoms | $V^0_{O6}$ defective 215 atoms |
|---|---|---|---|---|
| $Ti_1$-$Ti_2$ | 2.87 | 3.46 | 2.85 | 2.89 |
| $Ti_3$-$Ti_4$ | 2.99 | 3.17 | 3.50 | 2.99 |
| $Ti_5$-$Ti_6$ | 3.05 | 3.13 | 3.18 | 3.49 |
| $Ti_1$-$O_1$ | 1.84 | - | 1.88 | 1.84 |
| $Ti_2$-$O_1$ | 1.86 | - | 1.79 | 1.83 |
| $Ti_3$-$O_5$ | 1.96 | 2.10 | - | 2.11 |
| $Ti_4$-$O_5$ | 1.94 | 2.05 | - | 1.77 |
| $Ti_5$-$O_8$ | 2.00 | 2.05 | 2.06 | - |
| $Ti_6$-$O_8$ | 1.93 | 1.99 | 2.02 | - |

## IV. CONCLUSIONS

We apply the activation-relaxation technique nouveau coupled with the Quantum Espresso package to predict the +2 charged and the neutral vacancies diffusion barriers on the anatase (101) surface. Five inequivalent oxygen vacancy sites are selected to understand the surface and subsurface vacancy diffusion. For initial studies, the 72-atom anatase (101) surface slab is used. This small slab model stabilizes the topmost surface +2 charged vacancy due to surface reconstruction; hence subsurface-to-surface vacancy diffusion is observed. This is attributed to the small cell that is unable to accommodate the strong Coulomb repulsion between $Ti^{+4}$ ions and +2 charged vacancy. With sufficient surface relaxation, a larger 216-atom slab model show rather that both +2 charged and neutral vacancies diffusion in the same the top surface layer to the subsurface direction. The diffusion barriers for the +2 charged oxygen vacancy migration are 0.82 eV for direct diffusion, and 1.01 eV for inverse diffusion. The neutral oxygen vacancy diffusion shows a lower barrier for the direct diffusion to the subsurface i.e., 0.52 eV, and 0.81 eV inverse barrier for the surface migration. In addition, $V^{+2}_{O3}/V^{0}_{O3}$ surface vacancies can migrate toward the subsurface with competing diffusion mechanisms. In fact, $V^{+2}_{O3}$ to $V^{+2}_{O5}$ diffusion has a 0.73 eV barrier, which is smaller than the $V^{+2}_{O1}$ to $V^{+2}_{O5}$ activation barrier. As consequences of vacancy charged and competing diffusion mechanisms, a range of diffusion barriers for surface-to-subsurface oxygen vacancy migration on the anatase (101) surface is observed. Moreover, we investigate the bulk oxygen vacancy formation energies for both intrinsic vacancy types and their diffusion barriers. Bulk vacancies are less stable than the subsurface vacancies. The activation energy barrier for the +2 charged vacancy migration from the bulk to the subsurface is 0.19 eV, whereas neutral vacancy migration has to overcome the 0.27 eV barrier. Therefore, we unambiguously provide evidence for bulk to subsurface diffusion. Inverse barriers for bulk-to-subsurface diffusion suggest dynamic diffusion from bulk-to-subsurface and subsurface-to-bulk. Our results provide theoretical evidence for surface-to-subsurface and bulk-to-subsurface oxygen vacancy diffusion, and hence the experimentally observed accumulation of oxygen vacancies at the subsurface is well substantiated. Although our small surface slab stabilizes +2 charged vacancy on the topmost surface as opposed to more reliable larger slab model, it provides information might be useful to stabilize topmost surface oxygen vacancy. The topmost vacancy stability due to vertical displacement of $Ti^{+4}_{4c}$ ion neighboring the vacancy can be used to reverse the natural diffusion direction on the anatase (101) surface. Therefore, we suggest that such a surface reconstruction may be the origin of the reversibility of the diffusion direction from the subsurface to the surface. In addition, semi-local approximation and the Hubbard model computed vacancy formation energies and the diffusion barriers indicate that semi-local approximation is good enough to investigate diffusion barriers. Moreover, this study provides an in-depth understanding of the diffusion of intrinsic oxygen vacancies on the surface of the anatase (101) using the GGA-PBE and the DFT + $U$ levels of theories, as well as the diffusion barriers and diffusion paths for all inequivalent surface oxygen vacancy sites and the bulk, which are largely lacking in the literature. In addition, +2 charged and neutral vacancy diffusion is addressed in detail. More widely, this investigation sheds light also on the intricacies of modeling charged oxygen vacancy migration on the anatase (101) surface. Furthermore, this article may serve the experimental community in devising methods to engineer the (101) surface of anatase to acquire desired surface reactions by tuning the direction of diffusion through surface reconstruction.

## ACKNOWLEDGEMENTS

The authors would like to thank Prof. Annabella Selloni for initial discussions. J.D. is supported by the Institut de l'énergie Trottier' Scientific Director's research allocation. N.M. acknowledges partial support through a Discovery grant from the Natural Science and Engineering Research Council of Canada (NSERC). We are grateful to Calcul Québec and the Digital Research Alliance of Canada for generous allocation of computational resources.